\documentclass[11pt,twoside]{article}
\usepackage{./asp2014}

\aspSuppressVolSlug
\resetcounters

\begin{document}

\title{DEBCat: a catalogue of detached eclipsing binary stars}
\author{John~Southworth
\affil{Astrophysics Group, Keele University, Staffordshire, ST5 5BG, UK}
}

\paperauthor{John~Southworth}{astro.js@keele.ac.uk}{}{Keele University}{Astrophysics Group}{Newcastle-under-Lyme}{Staffordshire}{ST5 5BG}{UK}

\begin{abstract}
Detached eclipsing binary star systems are our primary source of measured physical properties of normal stars. I introduce DEBCat: a catalogue of detached eclipsing binaries with mass and radius measurements to the 2\% precision necessary to put useful constraints on theoretical models of stellar evolution. The catalogue was begun in 2006, as an update of the compilation by Andersen (1991). It now contains over 170 systems, and new results are added on appearance in the refereed literature. DEBCat is available at: \,\texttt{http://www.astro.keele.ac.uk/jkt/debcat/}
\end{abstract}

\section{History and motivation}

Detached eclipsing binary star systems (dEBs) are our primary source of measurements of the physical properties of stars. The masses, radii, and surface gravities of stars in dEBs can be measured empirically, and to a precision and accuracy of 1\% or better (e.g.\ Southworth et al., 2005b, 2007). With an effective temperature measurement we can obtain their luminosities directly. dEBs are the main checks and calibrators for theoretical stellar models (e.g.\ Pols et al., 1998), and thus form the foundation of stellar and galactic astrophysics.

Other important uses of dEBs include as direct distance indicators (Pietrzy\'nski et al.\ 2013; Southworth et al., 2005a), calibrators of asteroseismic scaling relations (Frandsen et al., 2013), probes of the chemical evolution of massive stars (Pavlovski et al., 2009), tracers of binary evolutionary processes (Maxted et al., 2013), and characterisation of the host stars of transiting extrasolar planets (Southworth, 2009, 2011).

The study of dEBs has a long history (see Stebbins, 1911; Russell, 1912). Catalogues of well-studied systems have been published by Popper (1967), Popper (1980), Harmanec (1988), Andersen (1991) and more recently by Torres et al.\ (2010).

\articlefigure{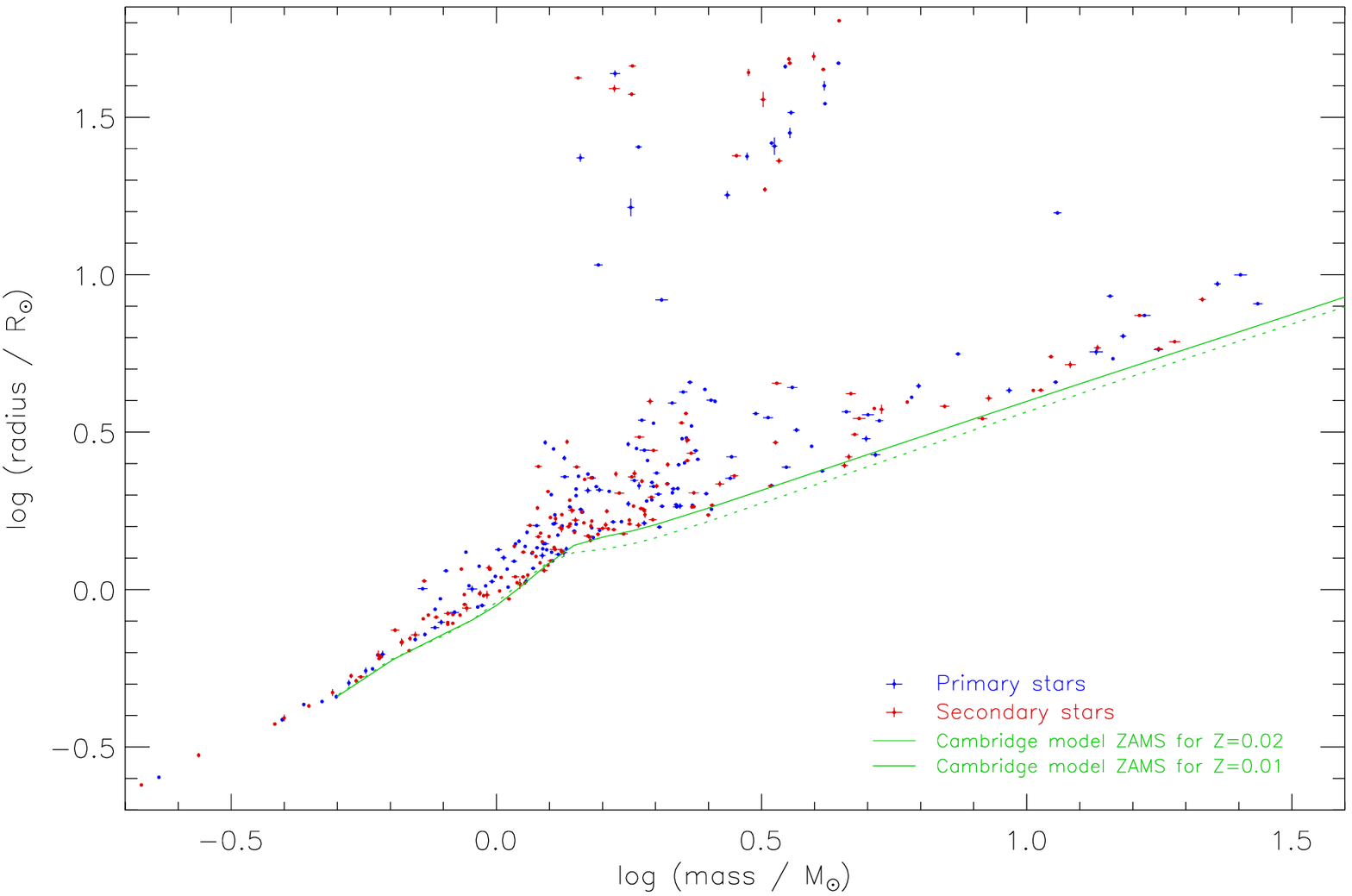}{fig1}{Plot of the masses and radii of the objects in DEBCat. The zero-age main sequences are from the Cambridge models (Pols et al., 1998) for metal abundances $Z=0.02$ and $0.01$.}


In early 2006 I constructed a catalogue of well-studied dEBs by updating the list of objects compiled by Andersen (1991). From this point onwards new results from refereed journals have been added as and when they are published. The result is \mbox{DEBCat}, which is available at: \,\texttt{http://www.astro.keele.ac.uk/jkt/debcat/}. A changelog has been kept since February 2008. The total number of objects in \mbox{DEBCat} is currently 171. For comparison, Torres et al.\ (2010) list 94 dEBs.

The requirements for inclusion in DEBCat are: (1) the evolution of the system has not been significantly affected by binarity, which restricts us to detached binary systems; (2) measurements of the effective temperatures of both stars are available; (3) the masses and radii of both components have been measured directly, i.e.\ without significant input from stellar theory, and to an accuracy of 2\%. This restricts us almost entirely to double-lined dEBs. The 2\% limit is relaxed for a few interesting objects.

In future I will continue to maintain DEBCat, add measurements of the orbital eccentricity and apsidal motion period, lodge a version with the {\it Centre de Donn\'es astronomiques de Strasbourg} (CDS),
and redesign the website (currently created from {\tt ascii} files by a {\sc fortran77} code). Current and future space missions, such as {\it Kepler}, K2, TESS and PLATO, should result in a huge increase in the number of well-studied dEBs suitable for inclusion in DEBCat.



\end{document}